\documentclass[prl,superscriptaddress,showpacs,twocolumn]{revtex4}
\usepackage{epsfig}

\begin{document}
\title{\bf  Colossal Effects in Transition Metal Oxides \\
can be caused by  Intrinsic Inhomogeneities}
\author{J. Burgy}
\affiliation{National High Magnetic Field Lab,
Florida State University, Tallahassee, FL 32306, USA}
\author{M. Mayr}
\affiliation{National High Magnetic Field Lab,
Florida State University, Tallahassee, FL 32306, USA}
\author{V. Martin-Mayor}
\affiliation{Dipartimento di Fisica, Universit\`a di Roma ``La Sapienza'',
Piazzale Aldo Moro 2, 00185 Roma, Italy}
\author{A. Moreo}
\affiliation{National High Magnetic Field Lab,
Florida State University, Tallahassee, FL 32306, USA}
\author{E. Dagotto}
\affiliation{National High Magnetic Field Lab,
Florida State University, Tallahassee, FL 32306, USA}

\date{\today}

\begin{abstract}
The influence of quenched disorder on the competition between ordered 
states separated by a first-order transition is investigated.
A phase diagram with features resembling quantum-critical behavior 
is observed, even using classical models. 
The low-temperature paramagnetic regime consists of 
coexisting ordered clusters, with randomnly oriented order 
parameters. Extended to manganites, this state is argued to have a
colossal magnetoresistance effect. A scale $\rm T^*$
for cluster formation is discussed. This is the analog of the
Griffiths temperature, but for the case of two 
competing orders, producing a strong susceptibility to external fields. 
Cuprates may have similar 
features, compatible with the large proximity effect 
of the very underdoped regime.
\end{abstract}
\maketitle

%
%

Complex phenomena such as
``colossal'' magnetoresistance (CMR) in manganites
and high temperature superconductivity (HTS) in cuprates
have challenged our understanding of correlated electrons~\cite{tokura}.
Recent developments unveiled a previously mostly ignored aspect of
doped transition-metal-oxides (TMO): these systems are 
{\it intrinsically} {\it inhomogeneous}, 
even in the best crystals.
{\bf (i)} 
The evidence in the CMR context is overwhelming. 
Experiments and theory provide a picture where 
competing ferromagnetic (FM) and charge-ordered (CO) states form
microscopic and/or mesoscopic coexisting clusters~\cite{cheong,review}. 
Exciting recent experiments~\cite{tokura2} 
identified features referred to as a ``quantum critical
point''
(QCP)~\cite{subir}  -- defined as the drastic reduction of ordering
temperatures near the zero temperature (T=0) transition between 
ordered states -- by modifying the A-site cation mean-radius 
$\langle$$r_A$$\rangle$ by chemical
substitution at fixed hole density (left inset of Fig.~1). 
The paramagnetic state in 
the QCP region -- where the Curie temperature $\rm T_C$ is the lowest --
is crucial to understand CMR phenomenology,
producing the largest CMR ratio~\cite{tokura,review,cheong}. 
{\bf (ii)}
In the HTS context, 
scanning tunneling microscopy (STM) studies of superconducting (SC) Bi2212 
revealed a complex surface with nm-size 
coexisting clusters~\cite{STM}. 
Underdoped cuprates also appear to be inhomogeneous~\cite{inhomo}.
In addition, a ``colossal'' proximity effect (CPE) was
reported on underdoped $\rm Y Ba_2 Cu_3 O_{6+x}$ over large
distances~\cite{decca}.

\begin{figure}
\epsfig{figure=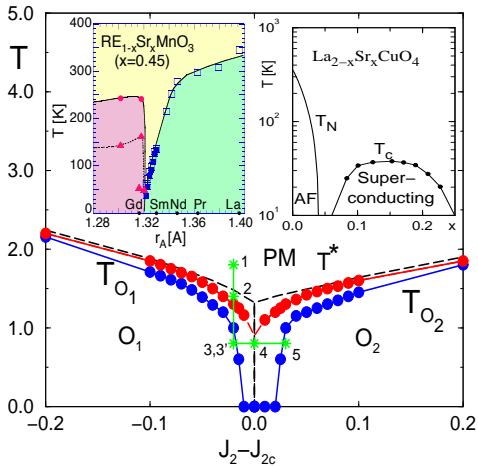,height=6.5cm}
\caption{
Phase diagram of the 2D $\rm J_1$-$\rm J_2$-$\rm J_4$ toy model
used to analyze the competition between ordered states. 
Results on 32$^3$ lattices (not 
shown) lead to a qualitatively similar phase diagram. $\rm J_1$=1
is the scale, and $\rm J_4$ is fixed to 0.2.
Disorder is incorporated such that $\rm J_2({\bf i j})$ at a link 
joining sites $\bf i$ and ${\bf j}$ is uniformly distributed between
$\rm J_2$-W/2 and $\rm J_2$+W/2. Blue (red) curve corresponds to
W=1.5 (0.75). Dashed black lines are the result without disorder
W=0, and T$^*$ denotes the clean-limit transition.
A Metropolis algorithm was used, on up to 512$^2$ lattices, calculating 
(i) the largest ordered cluster size [J. Hoshen {\it et al.},
Phys. Rev. B{\bf 14}, 3438 (1976)]
 and (ii) the 
order parameters for AF and collinear phases with a
spin structure factor maximized at momenta $(\pi,\pi)$ and $(\pi,0)$-$(0,\pi)$,
respectively. Ten or more realizations of disorder were used, found to
be sufficient for large systems. Insets are the
phase diagrams of Mn-oxides in the FM-CO competition 
region~\cite{tokura2}, and of
the single-layer Cu-oxide in standard notation
[J. B. Torrance {\it et al.}, Phys. Rev. B{\bf 40}, 8872 (1989)].
Points 1-5 are explained in Fig.~2.
}
\end{figure}

In this paper, 
the competition between two ordered states in the presence of quenched
disorder is investigated. These states are assumed 
sufficiently ``different'' that their low-T transition 
in the clean limit has 
{\it first-order} characteristics. The approach has similarities
with the classical work of Imry and Ma \cite{imry}. From 
the general considerations, doped TMOs are here considered, 
with intrinsic disorder caused by chemical 
substitution. For Mn-oxides,
a possible rationalization of the CMR effect is discussed,
with predictions including a scale $\rm T^*$ for
cluster formation -- the analog of the Griffiths temperature~\cite{griffiths} 
but in the regime of competing orders. For underdoped Cu-oxides,
a similar inhomogeneous picture is proposed.
The calculations are mainly carried out using a two dimensional (2D)
toy model of Ising spins, but similar data in three dimensions (3D) and for the
one-orbital manganite model have also been gathered.
Then, our conclusions appear
valid for a variety of models with competing orders. 
The actual Hamiltonian employed here, defined on a square/cubic 
lattice (spacing a=1) and with Ising variables, is
$\rm H$=
$\rm J_1 \sum_{\langle {\bf ij} \rangle} {{{ S^z}_{\bf i}}{{ S^z}_{\bf j}}}$ +
$\rm J_2 \sum_{\langle {\bf im} \rangle} {{{ S^z}_{\bf i}}{{ S^z}_{\bf m}}}$ +
$\rm J_4 \sum_{\langle {\bf in} \rangle} {{{ S^z}_{\bf i}}{{ S^z}_{\bf n}}}$,
in a standard notation.
Sites $\langle {\bf i}{\bf j} \rangle$ 
are at distance 1 (usual nearest-neighbors), 
$\langle {\bf i}{\bf m} \rangle$ at distance $\sqrt{2}$, and
$\langle {\bf i}{\bf n} \rangle$ at distance $\sqrt{5}$. 
The three couplings are
antiferromagnetic (AF). More than one coupling is 
needed to generate two competing T=0 states, and $\rm J_1$ and $\rm 
J_2$ are the natural ones. However, the clean-limit
first-order transition between those states was found to be more robust
if a small $\rm J_4$$\sim$0.2$\rm J_1$ coupling is added. 
The resulting competing states $\rm O_1$ and $\rm O_2$
are an AF state for low $\rm J_2$/$\rm J_1$, 
and a ``collinear'' AF state with rows (or columns) of spins up and down
for large $\rm J_2$/$\rm J_1$~\cite{landau}.
The main features of the toy model phase diagram are
common to a variety of models with competing tendencies.

The toy model phase diagram, without
disorder, is shown in Fig.~1, and it has the expected shape:
the ordering temperatures decrease and meet
at the clean-limit critical coupling 
$\rm J_{2c}$=0.7$\rm J_1$, and the low-T transition was found to be 
clearly first-order. The most interesting result in Fig.~1 is the influence
of disorder on the clean-limit diagram. 
The first-order transitions become continuous
with sufficiently large disorder, in agreement 
with previous work~\cite{cardy}. Critical
temperatures far from $\rm J_{2c}$ are not affected much  by the  
disorder strengths considered. However,  a {\it drastic}
reduction is observed near $\rm J_{2c}$. In fact, 
the Monte Carlo (MC) results suggest that the obtained
phase diagram is similar to the insets Fig.~1 for Mn- and Cu-oxides.
With increasing
disorder strength W, either a first-order line separating 
the competing phases
still survives at $\rm J_{2c}$ (red points), 
as in manganites,
or a disordered region of finite $\rm J_2$ width opens at
T=0 (blue points), as in single-layer cuprates. 
Note that the ordering temperatures exactly meet at T=0
for only one fine-tuned W. However, the overall shape of the
phase diagram resembles QCP behavior in a robust 
range of W. For this reason, our results are qualitatively described
as inducing ``quantum-critical-point like'' behavior, not a rigorous
expression but hopefully descriptive enough to be useful.

Sufficiently strong
quenched disorder will smooth first-order transitions~\cite{cardy}.
At $\rm J_{2c}$, this should occur at
infinitesimal W in 2D ~\cite{imry}, while a finite W is needed in 3D. Then,
at $\rm J_{2c}$ and with finite temperature, a paramagnetic state must 
be generated with growing disorder. 
Note also that our toy model is classical, 
but QCP-like behavior is nevertheless generated \cite{mccoy}.

\begin{figure}
\epsfig{figure=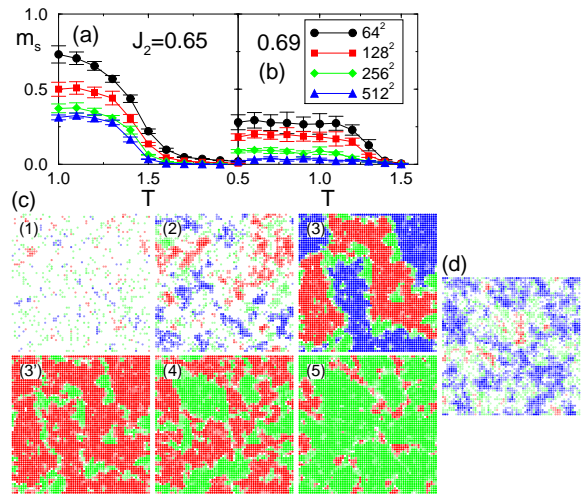,height=6.5cm}
\caption{
AF order parameter vs T for the toy model at fixed
W=1.5, using several lattice sizes with periodic boundary
conditions. (a) corresponds to $\rm J_2$=0.65 and
(b) to $\rm J_2$=0.69. Note in the latter the order parameter rapid
suppression as the size grows.
(c) Typical spin configurations representative of dominant
2D states. Shown are averages over 10 measurements, in
about 100 MC sweeps to avoid correlations, after thermalizing with
thousands of sweeps. Very similar results were obtained in 3D simulations.
{\bf (1,2,3)} are 
at $\rm J_2$=0.68, and 
T=2.00, 1.45 (near the resistance peak, see Fig.~3), 
and 0.80, respectively (see Fig.~1).
Green regions have collinear order, while red and blue indicate
N\'eel and ``anti-N\'eel'' order. The last two differ in the 
staggered order parameter sign, i.e. they intuitively are
$\uparrow \downarrow \uparrow \downarrow ...$ and
$\downarrow \uparrow \downarrow \uparrow ...$. The white
does not have a dominant order after
the MC sweeps considered here. Green/red/blue pale regions have weak order. 
{\bf (3$'$,4,5)} corresponds to T=0.8 and $\rm J_2$=0.68, 0.70 and 0.73,
respectively (see Fig.~1), and
the N\'eel and anti-N\'eel states are here given the same color (red),
while green remains collinear.
(d) Typical spin configuration at staggered field
$\rm H_s$=0.01, $\rm J_2$=0.68,
W=1.5, and T=1.45.
}
\end{figure}

Figure 1 is the result of a systematic computational effort.
As example, 
in Fig.~2a-b, the AF order parameter vs T is shown for 
$\rm J_2$ values outside and inside the coupling range where a T=0
disordered regime is obtained. For $\rm J_2$=0.69 note the 
order-parameter cancellation with increasing size (this coupling is not
critical at T$\neq$0).
Representative spin configurations are in Fig.~2c. Keeping $\rm J_2$
constant and varying T, three regimes are observed: {\bf (1)} A high-T regime,
where the system is disordered after MC time averaging. {\bf (2)} 
An intermediate
region  $\rm T_{O_1}$$<$T$<$$\rm T^*$ with preformed clusters, but with
uncorrelated order parameters giving a globally {\it paramagnetic}
state, similar to the Griffiths phase.
{\bf (3)} A low-T regime where the clusters from (2) grow 
in size, although the disorder is {\it uncorrelated} 
from link to link, and 
percolate upon cooling. Note 
that clusters with different signs for the order parameter are
separated by thin regions of the competing phase, providing a possible
mechanism for stabilizing domain walls.
Considering now fixed low-T but changing
$\rm J_2$, configurations (3$'$,4,5) are obtained. In this case, just the two
ordered phases are in competition (no white regions), 
and the transition between phases
appears percolative-like.

\begin{figure}
\epsfig{figure=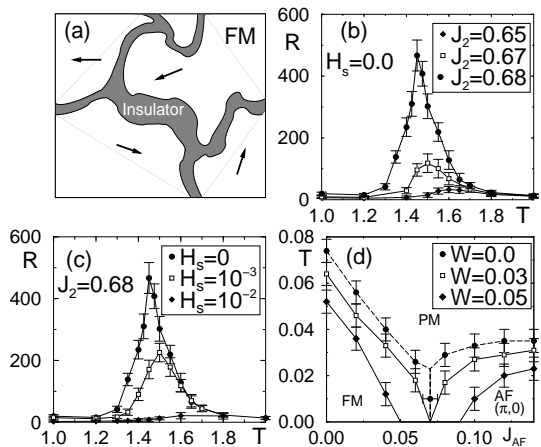,height=6cm}
\caption{
(a) Proposed
state for Mn-oxides in the CMR regime. 
(b) Resistance of the toy model after the equivalence
to manganite states is used (see text), using a 256$^2$ lattice, W=1.5,
at the couplings indicated.
The calculation is carried 
out by transforming a spin configuration
into a resistor network, with nodes centered at plaquettes
(\# nodes = 1/4 \# sites) and resistors between them. 
The values of the conductances of these resistors were established using
BB=RR=1.0, BR=0, WW=0.3, GG=BG=RG=WG=0,
$\rm WB$$=$$\rm WR$$=$0.5, 
where B, R, G, and W stand for blue, red,
green and white regions (see Fig.~2c),  AA$'$= $\alpha$ means that
the resistor between plaquettes in the A (=B,R,G,W) 
and A$'$ states has value
$\alpha$ (arbitrary units), and AA$'$=A$'$A. Other values for WW 
and WB lead to similar results, and BB defines the scale.
Note that the conductivity should
be {\it spin dependent}, 
and a BR link (when an
electron moves from a spin-up to a spin-down region) has zero conductance.
The algorithm used to obtain the total conductance 
is exact~\cite{frank}.
(c) Resistance (arbitrary units) vs T, at external 
fields $\rm H_s$ indicated, using a $256^2$ lattice, $\rm J_2$=0.68,
and W=1.5. 
(d) $\rm T^*_C$ 
results for the one-orbital manganite model using $8^2$ and
$16^2$ clusters, density $x$=0.5, infinite Hund coupling, and hopping t=1. 
With disorder, the $\rm J_{AF}$ couplings are randomly distributed between
$\rm J_{AF}$-W and $\rm J_{AF}$+W (W indicated).
In practice, $\rm T^*_C$ was defined when the spin correlations at
distance $\sqrt{2}$
dropped below 40\% of the maximum value. 
}
\end{figure}

The main features of the
results Figs.~1-2 -- shape, clustered structure, the QCP-like
behavior -- are believed to be qualitatively general. In fact,
simulations of one-orbital models (below) and other models
studied in this effort 
give a similar phase
diagram. Of course, the analogy should not be taken too far, e.g.
critical exponents may not be universal since the Ising model
underlying symmetries are quite different from those of realistic systems.
However,
it is worth investigating the consequences of the general
phase diagram Fig.~1 for
materials where two states strongly compete, such as in Mn-oxides. In
this context, if a simulation of a realistic model with FM and AF phases
on a huge lattice were possible, FM and AF clusters analogous of
Fig.~2 would be found. Then, a reasonable way to bypass that 
(currently impossible) computational effort is to simply translate Fig.~2 into
manganite language. This is a speculation, 
but hopefully the essence of the problem is preserved by
the procedure. The proposed translation
links order $\rm O_1$ with ferromagnetism, 
with order parameters pointing in different directions for different
clusters, while $\rm O_2$ corresponds to charge-ordering. 
Translating Fig.~2c(2) into Mn-oxide
language leads schematically to Fig.~3a, our proposed CMR state. 
The preformed FM clusters
have uncorrelated moment orientations, and zero global
magnetization. 
Note also that the ``depth'' of the
QCP-like feature is not universal, 
it depends on the disorder strength.

To test the relevance of Fig.~3a to CMR manganites, 
a resistor network calculation was set up. 
Translating to Mn-oxide language, as explained before, the MC generated
configurations were mapped into a resistance grid (see
caption of Fig.~3b). For up (down) spins, 
the blue regions of Fig.~2c -- analog of positive
magnetization FM clusters -- have high (low)
conductivity, the red regions have low (high) conductivity, and
the green regions are insulating. The Kirchoff equations
were solved exactly, leading to the results Fig.~3b.
In agreement with intuition, the non-percolated state Fig.~2c(2)
has a {\it large resistance} for both spins up and down, while
the percolated low-T or
disordered high-T states have far better conductance. Note that the 
resistance peak intensity increases  as the ordering temperature is
reduced varying $\rm J_2$, analog of $\langle$$r_A$$\rangle$, closer 
to the QCP-like regime.

The rotation of the large moments of the
preformed FM clusters (Fig.~3a)
may occur with small magnetic fields. These
effects are mimicked in the toy model using a staggered
external field $\rm H_s$ which favors $\rm O_1$ clusters
with order parameter $\rm M_s$$>$0 
(blue, Fig.~2c)
to the detriment of $\rm O_1$ clusters with -$\rm M_s$ (red, Fig.~2c), or 
$\rm O_2$ regions. Figure~2d confirms the 
rapid generation of positive $\rm O_1$ order in
the region $\rm T_{O_1}$$<$T$<$$\rm T^*$ with
tiny fields 0.01$\rm J_1$. This
severely affects transport (Fig.~3c), i.e. a modest
field transforms the intermediate T cluster state
into a fairly uniform state with robust conductance. Results
Fig.~3c -- the main results of this paper --
are similar to those found in Mn-oxides, 
and a huge MR ratio $\rm [R(0)-R(H_s)]$/$\rm R(H_s)$ of 
$\sim$4$\times$10$^3$~\% was obtained at $\rm H_s$=10$^{-2}$ \cite{comment5}. 

In addition,
there are already experimental indications in Mn-oxides for the
existence of a temperature scale $\rm T^*$ for uncorrelated cluster 
formation~\cite{ibarra}, which should be ubiquitous in low-bandwidth 
manganites~\cite{pseudogap}. 

\begin{figure}
\epsfig{figure=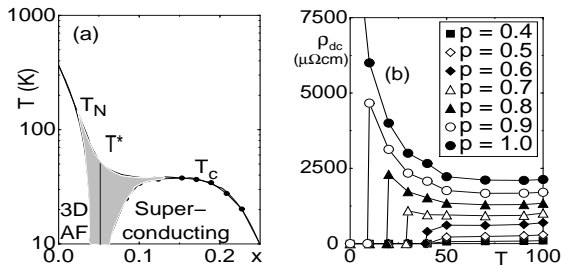,height=4cm}
\caption{
(a) Conjectured HTS phase diagram. 
Black lines should be the actual phase boundaries without
disorder. The shaded region is conjectured to have metallic (SC) and
insulating (AF) coexisting regions in the real materials.
(b) Resistivity $\rho_{ab}$ 
vs T, from a random-resistor network calculation as 
in Ref.\cite{mayr}, where details can be found. A 50$\times$50
cluster was used, with $\rho_{ab}$ for insulating (optimal doping)
fraction p=1.0 (0.0) 
taken from LSCO x=0.04 (0.15) data [H. Takagi {\it et al.},
Phys. Rev. Lett. {\bf 69}, 2975 (1992). See also
Y. Ando {\it et al.}, cond-mat/0104163]. The inset labels
are the p fractions at 100 K, all of which are smoothly reduced 
with decreasing T
until percolation to a SC state occurs at p=0.5.}
\end{figure}

The phase diagram Fig.~1 is representative
of more realistic models. Fig.~3d contains MC results
for the one-orbital model~\cite{victor}, in which
the coupling $\rm J_{AF}$ between localized spins is varied
 to induce a metal-insulator
transition~\cite{review}. 
Without disorder, the T$\sim$0 transition is known to be
first-order between FM and AF states, the latter
with rows or columns of spins up and down~\cite{review}.
The ``characteristic'' ordering temperatures $\rm T^*_C$,
at which spin correlations become robust upon cooling, are shown 
vs. $\rm J_{AF}$. Note the similarity with Fig.~1.

The results in Figs.~1-2 can also be adapted to cuprates.
The high-Tc phase diagram (inset Fig.~1) shows a suppression of AF and 
SC order in a region usually labeled ``spin-glass'', whose origin is
unclear.
Considering these diagrams together
with the CPE results~\cite{decca},
it is conjectured that the very underdoped cuprate 
state may not be homogeneous but results from a SC vs. {\it doped} AF
competition after disorder is considered. Inhomogeneities (clusters)
should be present even within ordered phases [Fig.~2c(3)]. 
Stripe states are candidates for the doped AF state \cite{emery}. 

The proposed clean-limit phase diagram is in Fig.~4a, with
 a vertical first-order transition line,
as cuprates have upon electron doping~\cite{harima}, and
heavy fermions with varying pressure. 
The shaded region may contain 
a mixture of stripe-like and preformed SC islands~\cite{iguchi}. 
Due to the general character of the discussion of Figs.~1-2,
{\it colossal effects should be ubiquitous when ordered phases compete},
and they could appear in cuprates as well.
CPE ~\cite{decca} could be a manifestation, with 
preformed SC clusters 
percolating under the influence of nearby SC materials. 
To further check this hypothesis, Fig.~4d contains results
of a phenomenological random-resistor calculation of resistivity 
vs T~\cite{mayr} in rough agreement with
experiments~\cite{ichikawa}.

Summarizing, the results presented in this paper \cite{burgy}
suggest
that ``colossal'' effects in TMO's could originate in intrinsic 
inhomogeneities. These large effects  may be more general
than previously anticipated. In our studies,
the analog of the classical 
Griffiths regime -- usually associated with weak effects -- 
is here much more robust, strongly susceptible to external fields.

This work was supported by NSF-DMR-9814350 and
the Computational Science
and Information Technology school at Florida State University.
The authors thank Y. Ando, D. N.  Argyriou, S. L. Cooper,
J. C. S. Davis, R. Decca, A. Feiguin, D. Gingold, N. Nagaosa, 
S. Sachdev, T. Senthil, P. Schiffer, and Y. Tokura for comments.

\end{document}